# Origin of enhanced dynamic nuclear polarization and all-optical nuclear magnetic resonance in GaAs quantum wells

G. Salis and D. D. Awschalom

Department of Physics, University of California, Santa Barbara CA 93106

Y. Ohno and H. Ohno

*Laboratory for Electronic Intelligent Systems, Research Institute of Electrical Communication, Tohoku University, 2-1-1 Katahira, Aoba-ku, Sendai 980-8577, Japan*

**Abstract**

Time-resolved optical measurements of electron-spin dynamics in a (110) GaAs quantum well are used to study the consequences of a strongly anisotropic electron g-tensor, and the origin of previously discovered all-optical nuclear magnetic resonance. All components of the g-tensor are measured, and a strong anisotropy even along the in-plane directions is found. The amplitudes of the spin signal allow the study of the spatial directions of the injected spin and its precession axis. Surprisingly efficient dynamic nuclear polarization in a geometry where the electron spins are injected almost transverse to the applied magnetic field is attributed to an enhanced non-precessing electron spin component. The small absolute value of the electron g-factor combined with efficient nuclear spin polarization leads to large nuclear fields that dominate electron spin precession at low temperatures. These effects allow for sensitive detection of all-optical nuclear magnetic resonance induced by periodically excited quantum-well electrons. The mechanism of previously observed $\Delta m = 2$ transitions is investigated and found to be attributable to electric quadrupole coupling, whereas $\Delta m = 1$ transitions show signatures of both quadrupole and electron-spin induced magnetic dipole coupling.

# I. INTRODUCTION

The coherent dynamics of electron spins in semiconductors is of fundamental interest for novel "spintronic" devices,[1] and can be investigated using time-resolved optical techniques. A well-known manifestation of spin coherence is the beating of time- and polarization-resolved photoluminescence due to precession of electron spins about an applied magnetic field.[2] Because this type of beating involves the spins of recombining charges, the observable duration is limited to the charge recombination time. Spin coherence detected using time-resolved absorption[3] or Faraday rotation[4] (FR) can in principle be monitored over arbitrary long time scales in doped semiconductors and heterostructures.[5] Using time-resolved FR, spin lifetimes exceeding the charge recombination times by four orders of magnitude have been observed in n-doped GaAs.[6] Besides enabling macroscopic transport of spin coherence[7] and potentially semiconductor-based quantum-computation devices,[8] such long spin lifetimes allow for precise measurement of the electron-spin precession frequency. Due to the contact hyperfine interaction with nuclear spins, time-resolved FR can be employed as sensitive magnetometer for nuclear spin polarizations ("Larmor magnetometry").[9,10,11] In such experiments, dynamic nuclear polarization (DNP) is used to achieve nuclear spin polarizations exceeding the thermal equilibrium values by orders of magnitude. DNP has been demonstrated in bulk semiconductors,[12,13,14] quantum wells (QWs),[15,16,17] and quantum dots.[18] Combining DNP and Larmor magnetometry enables optical detection of nuclear magnetic resonance (NMR) with direct and accurate measurement of nuclear spin polarization at arbitrary magnetic fields. This is in contrast to the conventional approach of optically detected NMR, which relies on cw measurements of photoluminescence polarization reduced by the Hanle effect.[14]

In NMR experiments, the resonance is usually induced by externally applied radio-frequency magnetic fields. It has been demonstrated in Ref. 19 that the radio-frequency tipping field can be replaced by modulated light interacting indirectly with the nuclei through the electron system. Such an "all-optical" NMR scheme has been realized recently in Refs. 9,10 using a mode-locked pump laser. In this approach, the resonance condition is obtained when the laser repetition rate matches the nuclear precession period in a static magnetic field. Resonant depolarization of the nuclear spin is observed through electron Larmor magnetometry. In conventional NMR experiments, the radio-frequency magnetic fields penetrate the whole semiconductor. Applying the all-optical scheme in a quantum-confined electronic structure[10] leads to spatial confinement of the nuclear excitation and therefore to unprecedented localization of all the NMR mechanisms – initial polarization, excitation, and detection.

Here, we discuss in more depth electron spin coherence and all-optical NMR measurements in a narrow GaAs QW grown on a GaAs (110) substrate. Electrons confined in such QWs feel fluctuating effective fields from spin-orbit coupling that are oriented along [110] to lowest order in perturbation theory.[20] This special symmetry significantly reduces D'yakonov-Perel spin scattering,[21] and spin lifetimes > 1 ns were measured at room temperature.[22] In order to characterize the electron-spin dynamics without contributions from nuclear effects, the samples are measured at intermediate temperatures (135 K), where DNP is inefficient. By varying the angle between the sample normal and the magnetic field **B**, we are able to measure the strongly anisotropic electron g tensor. Because the FR signal measures a projection of the electron spin, the relative orientation of the precession axis can be reconstructed, and we find an expected deviation of its direction from **B**. Furthermore the data suggests that electron spin is injected along the quantization axis of the QW, and not along the direction of light propagation. The g-factor anisotropy is found to be crucial for efficient nuclear spin polarization, leading to additional modifications of both electron-spin precession frequency and orientation of the precession axis. At liquid He temperatures, lowering the laser pump intensity reduces DNP, and contributions from the nuclear spin and the anisotropic electron g-tensor can be separated. This technique allows for a quantitative determination of the nuclear spin polarization.

Monitoring the nuclear spin polarization, we study resonant nuclear depolarization induced by a laser pulse train (all-optical NMR). In Ref. 10, such resonances were observed also at laser repetition rates matching twice the nuclear precession frequency, and were interpreted as $\Delta m = 2$ transitions within the nuclear spin states with angular quantum number $m$. Using two pump beams, we find that for $\Delta m = 2$, the depth of the resonance does not depend on the spin-polarization of the electrons periodically injected by the second pump beam. Together with the lack of observed sidebands in case of spin-modulation of the resonantly injected electrons, we rule out interaction with the electron spin as tipping mechanism for $\Delta m = 2$, in agreement with selection rules for magnetic dipole coupling. Without injection of electrons, the resonance disappears, suggesting that the charge of the injected carriers depolarizes nuclear spin through interaction with nuclear electric quadrupolar moments. On the other hand, $\Delta m = 1$ resonances are found to be deeper when mediated by spin-polarized electrons, and sidebands occur in case of spin-modulation. This suggests that the originally proposed hyperfine mechanism for nuclear spin tipping significantly contributes to all-optical NMR at $\Delta m = 1$.

The paper is organized as follows: In section II, we describe the experimental setup and the measured samples. Experimental characterization of the spin-coherence, including measurements of the anisotropic electron g-tensor at 135 K and at 5 K is discussed in section III. A model for DNP describes the measured angle-dependence of the electron-spin precession frequency at low temperatures. Results on all-optical NMR are presented in section IV together with studies of nuclear tipping mechanisms responsible for $\Delta m = 1$ and $\Delta m = 2$ transitions. We conclude in section V.

## II. EXPERIMENTAL SETUP

The samples studied are single, 7.5 nm wide GaAs QWs with $Al_{0.4}Ga_{0.6}As$ barriers on both sides. The samples are grown on (110) GaAs wafers by molecular beam epitaxy and are modulation doped with a nominal Si density of $4 \cdot 10^{11}$ cm$^{-2}$. The mobility and electron density at 300 K are 1700 cm$^2$V$^{-1}$s$^{-1}$ and $9 \cdot 10^{10}$ cm$^{-2}$, respectively, as measured in a Hall bar configuration.

The samples are glued on fused silica, and the GaAs substrate is removed by mechanical polishing and selective etching, allowing for measurement of the FR in transmission. They are placed in an optical cryostat with a variable temperature insert, where temperatures between 2 K and 300 K can be achieved. The energy of a 100 fs Ti:sapphire laser with a pulse repetition rate of 76 MHz is tuned close to the heavy-hole absorption edge of the QW (1.572 eV at 5 K, 1.537 eV at 135 K). Pump and probe pulses are separated with a beam splitter. A delay line allows for a tunable time delay $\Delta t$ between pump and probe pulses ranging from –0.3 to 2.7 ns. The circularly polarized pump pulses and the linearly polarized probe pulses are focused to spatially overlap on the sample surface within a focal spot diameter of ~50 µm. Using an average pump power of 1.2 mW leads to injection of around $10^{10}$ cm$^{-2}$ electron-hole pairs per pulse, well below the doping density. The angle between the pump and probe beam after passing through the lens is less than 5°. We measure the FR, i.e. the rotation angle of the transmitted probe beam's linear polarization, using a balanced photo-diode bridge. The pump beam is chopped at frequencies between 1 and 6 kHz, allowing for lock-in detection of the FR, which is proportional to the carrier's spin component along the probe beam. Electrons and holes recombine on a time scale of 100 ps, as inferred from measured time-resolved absorption. Thereafter, a spin imprint in the doping electrons is detected, and varying $\Delta t$ reveals the precessional relaxation of such spins. We measure transverse spin lifetimes around 10 ns at low magnetic fields and 5 K.

Figure 1(a) shows the measurement geometry: A magnetic field **B** of up to 7.5 T is applied perpendicular to both pump and probe beams. The sample can be rotated about an axis perpendicular to **B** and the laser beams. The rotation angle $\alpha$ is measured between **B** and the sample's in-plane direction, which is either the $[\bar{1}10]$ or [001] direction, depending on how the sample is mounted.

## III. ELECTRON SPIN COHERENCE

Due to quantum-confinement and penetration of the electron wave function into the AlGaAs barriers, the g-factor in a narrow QW can differ substantially from the bulk

GaAs value[23,24]. Heavy- and light-hole splitting in a QW leads to different values of *g* for **B** oriented in-plane or along the quantization axis[24,25,26,27]. Generally, the g-factor can be expressed as a tensor $\hat{\mathbf{g}}$. Due to crystal symmetry and the direction of the QW, the main axes of $\hat{\mathbf{g}}$ are expected to be the growth axis [110] (*z*), the [001]-direction (*y*) and the [$\bar{1}$10]-direction (*x*). With our high-resolution measurement of the electron-spin precession frequency, $\Omega_L$, we are able to determine all three components of $\hat{\mathbf{g}}$, and find strong anisotropy even for the two in-plane directions[28]. The anisotropy gives rise to a dependence of $\Omega_L$ on the orientation of **B** with respect to the coordinate system of the sample. Furthermore, the precession axis $\mathbf{\Omega} = \hat{\mathbf{g}}\mathbf{B}\mu_B/\hbar$ differs from the direction of **B**, which can be observed in the FR measurements and has important consequences for the nuclear polarization at lower temperatures.

In order to identify g-factor anisotropies in the absence of nuclear polarization, we first discuss measurements at 135 K, where DNP is inefficient.[10] Figure 1(c) shows FR measurements at *B* = 4 T with **B** oriented in the (*x*, *z*)-plane. For α = 0° (**B** || $\hat{\mathbf{x}}$), the signal oscillates about a small constant offset and is well fit by an exponentially decaying oscillation with an angular frequency of $\Omega_L$ = 13.3 GHz, corresponding to a g-factor along [$\bar{1}$10] of $|g_x|$ = 0.0376. For α = 25°, the oscillating signal is superimposed on a non-oscillating and exponentially decaying background. This background is due to a non-precessing electron-spin component $\mathbf{S}_\parallel$ [Fig. 1(b)], which has a component along the probe-beam direction, as explained below. At this angle, we find $\Omega_L$ = 24.4 GHz. For orientation of **B** in the (*x*, *z*)-plane, the effective g-factor is given by

$$g = \sqrt{g_x^2 \cos^2\alpha + g_z^2 \sin^2\alpha}, \tag{1}$$

where $g_x$ and $g_z$ are the respective components of the g tensor. The solid circles in Fig. 2(a) show $g$ measured as a function of $\alpha$ between -30° and 30°. A fit to the above equation yields $|g_x| = 0.0376 \pm 0.0002$ and $|g_z| = 0.1415 \pm 0.0005$. Mounting the sample with **B** in the $(y, z)$-plane allows the measurement of the tensor component $g_y$. The open squares in Fig. 2(a) show the $\alpha$-dependence of $g$ in this case. This dependence is described by a modified Eq. (1), where $g_x$ is replaced by $g_y$. A fit to the data yields $|g_y| = 0.0184 \pm 0.0004$ and $|g_z| = 0.1423 \pm 0.0006$. The two geometries yield the same value for $g_z$ within the error bars, which reflect only the quality of the fit. Systematic errors from field- and time-delay calibration are estimated to be on the order of 1 percent. Another error arises from possible misalignment of the sample's in-plane directions with respect to the rotation axis. For this angle, we estimate a precision of ±10°, which gives rise to the errors included in Tab. I, which summarizes the components of the g-tensor at 135 K. The value for $|g_y|$ is more than a factor of two smaller than $|g_x|$ and almost eight times smaller than $|g_z|$. Because the $\alpha$-dependence of $g$ is given by the squares of the g-tensor components, we cannot determine the sign of the g-tensor components with this technique.

The precession axis **Ω** is tilted away from the $x$ or $y$ axis by an angle $\gamma$, given by $\tan\gamma = g_z/g_{x,y} \tan\alpha$ [Fig. 1(b)]. For an anisotropic g-tensor, **Ω** is generally not collinear with **B**, and the electron spin **S** contains both a precessing ($S_\perp$) and a non-precessing ($S_\parallel$) component[29], as shown in Fig.1(b). We fit the FR data to the sum of $a_1 \exp(-\Delta t / T_1)$ and $a_2 \exp(-\Delta t / T_2) \cos \Omega_L t$. The two parameters, $T_1$ and $T_2$, are effective longitudinal and transverse spin lifetimes with respect to the precession axis. If one assumes that the FR measures the projection of **S** onto the probe beam direction, the amplitudes $a_1$ and $a_2$ are proportional to the respective projections of $\mathbf{S}_\parallel$ and $\mathbf{S}_\perp$, respectively.

The symbols in Fig. 2(b) show the obtained amplitude $a_1$ of the non-oscillating FR component, normalized by $a_1 + a_2 = 1$. Squares are for **B** in the (*y*, *z*)-plane, whereas the circles are for **B** in the (*x*, *z*)-plane. These amplitudes reveal the direction of **Ω**, and can be compared to those calculated from the known g-tensor components. Using Snell's law $n \sin \beta = \sin \alpha$, we calculate the angle β between the pump beam inside the sample and [110] [Fig. 1(b)]. Assuming that **S** is initially oriented along the pump-beam direction, we find $a_1 = S \sin^2(\gamma - \beta)$ and $a_2 = S \cos^2(\gamma - \beta)$. The amplitudes are proportional to the amount of injected spin, *S*, and depend on the relative sign of $g_z$ and $g_{x,y}$ through the angle γ. The solid lines in Fig. 1(c) show calculated $a_1$ assuming an index of refraction $n = 3.5$ (corresponding to the GaAs value) and same signs for the g-tensor components. The calculated values are consistently smaller than the measured ones. If we assume opposite signs for $g_z$ and $g_{x,y}$ [dotted lines in Fig. 2(b)], the calculated values are larger than the measured ones. This deviation can be explained by instead assuming that spin is injected along the sample normal, i.e. along the [110] direction. Such a scenario is motivated by the fact that the pump pulses couple predominantly to heavy hole states, which are split off from the light holes in a QW. This leads to a preferred initial orientation of the electron spin along the growth direction of the QW[30]. In Fig. 2(b), the dashed lines show calculated values for $a_1$ assuming spin-injection along [110].

Furthermore, it was assumed that the FR is proportional to the spin component along the sample normal. The values calculated with this model fit the measured data much better, suggesting that indeed the injected spin component is oriented closer to the sample normal than to the refracted pump-beam direction (assuming $n = 3.5$). Because $a_1 \propto \sin^2 \gamma$ in case of spin injection along [110], the amplitude does not depend on the signs of the g-tensor elements, and no information about these signs can be inferred. It has been shown both theoretically and experimentally in Refs. 24,25 that in a (001) GaAs QW, the electron g-factor along the quantization axis is smaller than the in-plane components. If we assume that this is also true for a (110) QW, the sign of $g_z$ can be determined from the absolute values of the g-tensor components. Because we find $|g_x|, |g_y| < |g_z|$, $g_z$ must be negative at 135 K, while the signs of the in-plane components can either be positive or negative.

At lower temperatures, the pump pulses dynamically polarize the nuclear spins over time-scales of minutes (corresponding to the long spin-lattice relaxation times of nuclei). Such nuclear polarization occurs both for injecting electron spin perpendicular to **B**, and for tilted samples with a spin component parallel to **B**. In both cases, the pump-induced polarization of electron spins differs from thermal equilibrium, leading to hyperfine-induced exchange of angular momentum with the nuclear spins and therefore to a hyperpolarization of the nuclei. However, polarization of electron spin collinear to **B** usually leads to an electron spin population further out of equilibrium and therefore to a larger nuclear spin polarization.[12] By tilting the sample by an angle α [Fig. 1(b)], a component of $\mathbf{S}_\parallel$ along **B** is introduced, which dominates DNP even for small α. We find that DNP sensitively depends on the in-plane orientation of **B**, which reflects the dependence of $S_\parallel$ on the anisotropic electron g-factor.

The nuclear spin polarization modifies coherent electron spin dynamics due to the contact hyperfine interaction, which can be described by an effective nuclear field $\mathbf{B_n}$ acting on the electron spin. If $\mathbf{B_n}$ has a component perpendicular to the injected electron spin, **S**, it leads to a change in the electron spin precession frequency, which we measure using time-resolved FR. Fig. 3(a) shows the apparent g-factors ($\hbar\Omega_L/\mu_B B$) vs. $\alpha$ measured at 5 K and $B = 4$ T oriented in the $(x, z)$-plane. The pump pulses are either right- (open diamonds) or left-circularly (filled diamonds) polarized with an average intensity of 0.92 mW. The data displays a pronounced asymmetry with the sign of $\alpha$. The angle-dependence is inverted when the helicity of circular polarization is reversed. The same inversion is observed when the sign of $B$ is changed[10]. This agrees with the picture of DNP along the projection of $S_\parallel$ on **B**, which reverses sign when $\alpha$ or **B** cross zero. In addition, the $B$-dependence of $\Omega_L$ deviates substantially from a linear increase[10], revealing the non-Zeeman-like contribution to the precession frequency.

We obtain the real g factor by reducing the laser pump intensity, which decreases DNP. The squares in Fig. 3(a) show additional data for right-circularly polarized pump pulses at 90 μW. The asymmetry is strongly reduced, and we can fit the $\alpha$-dependence with the model of an anisotropic g-tensor in order to obtain the g-tensor components at 5 K. The fit (solid line in Fig. 2) yields $|g_x| = 0.031 \pm 0.005$ and $|g_z| = 0.17 \pm 0.01$. The difference between the symbols and the solid line is due to DNP. For the other in-plane direction, we obtain $|g_y| = 0.041 \pm 0.005$ and $|g_z| = 0.15 \pm 0.02$ [Fig. 3(b)]. For both in-plane orientations, the nuclear polarization changes sign at $\alpha = 0$. However, the angle-dependence is distinctively different for the two geometries. For **B** in the ($x$, $z$)-plane, the nuclear polarization is peaked at $\alpha = -6°$ and $4°$ and disappears around $\pm 20°$. For the other geometry [Fig. 3(b)], nuclear polarization steadily increases for negative $\alpha$, and remains about constant for positive $\alpha$ after a sharp increase. These differences are mainly due to the in-plane asymmetry of the g-tensor, leading to different $\mathbf{S}_\parallel$ in the two cases. In addition to this, there might be an intrinsic asymmetry of the hyperfine-interaction, responsible for different DNP efficiencies for the two orientations.

In the following, we calculate the angle-dependence of the electron-spin precession frequency including DNP. Such an investigation allows determining the sign of the components of $\hat{\mathbf{g}}$ [31]. If the applied field **B** is bigger than both the internuclear dipolar field (on the order of a few Gauss) and the hyperfine magnetic field from polarized electron spins, the average nuclear spin $\langle \mathbf{I} \rangle$ can be written as[14]

$$\langle \mathbf{I} \rangle = \frac{4}{3} I(I+1) f \frac{\langle \mathbf{S} \rangle \cdot \mathbf{B}}{|\mathbf{B}|^2} \mathbf{B}, \qquad (2)$$

where $\langle \mathbf{S} \rangle$ is the time-averaged electron spin, $I = 3/2$ the spin of the nuclei, and $f$ a leakage factor describing electron-spin relaxation other than through hyperfine-induced flipping of nuclear spins. Equation (2) is derived for $|\langle \mathbf{I} \rangle| \ll I$. For simplicity, we assume that **S** initially points along the $z$-direction, is of magnitude ½, and precesses about **Ω**. Generally, **S** can be decomposed into $\mathbf{S}_\parallel$ and $\mathbf{S}_\perp$ [Fig. 1(b)]. The transverse component $\langle \mathbf{S}_\perp \rangle$ is reduced due to spin precession, and we find that even for small $\alpha$, $|\langle \mathbf{S}_\perp \rangle| \ll |\langle \mathbf{S}_\parallel \rangle|$ and therefore $\langle \mathbf{S} \rangle \approx \langle \mathbf{S}_\| \rangle = c\mathbf{S}_\|$, where $c < 1$ reflects the exponential decay of $\mathbf{S}_\|$. The component $\mathbf{S}_\|$ is given by the projection of $\mathbf{S}$ on $\mathbf{\Omega}$, therefore

$$\langle \mathbf{S} \rangle \approx c \frac{\mathbf{S} \cdot \mathbf{\Omega}}{|\mathbf{\Omega}|^2} \mathbf{\Omega}. \tag{3}$$

The total energy of an electron spin consists of the Zeeman energy $\mu_B \hat{\mathbf{g}} \mathbf{B} \cdot \mathbf{S}$, and the hyperfine energy $A\langle \mathbf{I} \rangle \cdot \mathbf{S}$. This means that electron spin precesses about an axis given by

$$\mathbf{\Omega}_{tot} = \hat{\mathbf{g}} \mathbf{B} \mu_B / \hbar + A\langle \mathbf{I} \rangle / \hbar = \mathbf{\Omega} + \mathbf{\Omega}_\mathbf{n}. \tag{4}$$

According to Ref. 13, the constant $A/\hbar$ amounts to 137 GHz for GaAs. Inserting Eq. (2) and (3) into Eq.(4) provides the electron-spin precession frequency $\Omega_L = |\mathbf{\Omega}_{tot}|$. A self-consistent effect is neglected in this derivation because $\langle \mathbf{I} \rangle$ is calculated using the component of $\mathbf{S}$ along $\mathbf{\Omega}$. However, $\mathbf{S}$ precesses about $\mathbf{\Omega}_{tot}$, which can only be calculated if one already knows $\langle \mathbf{I} \rangle$. We neglect this self-consistency in the following discussion.

In Fig. 4, we compare the experimental data shown in Fig. 3 with Eq. (4). The only fit parameters are the signs of the g-tensor components, $f$, and $c$. We combine the latter two into a parameter $\xi = 5f \cdot c/3$ which indicates the percentage of nuclear polarization when $\mathbf{S}_\|$ is collinear to $\mathbf{B}$. The open diamonds in Fig. 4 show data of $\Omega_L$ measured with right circularly polarized pump pulses of average intensity 0.92 mW. For $\mathbf{B}$ applied in the $(x, z)$-plane, we obtain best agreement with $\xi$ between 0.12 and 0.15, and opposite signs for $g_x$ and $g_z$ [solid lines in Fig. 4(a)]. The calculated data reproduces the two peaks with different signs at positive and negative angles, as well as the disappearance of nuclear polarization at $\alpha \approx \pm 20°$. The reason for this disappearance is the rotation of $\mathbf{\Omega}$, and therefore $\langle \mathbf{S} \rangle$, into a direction perpendicular to $\mathbf{B}$, leading to inefficient DNP according to Eq. (2). If same signs are chosen (dotted line), the nuclear polarization does not disappear at $\alpha \approx \pm 20°$, and the data cannot be reproduced at all. Figure 4(b) shows data for the $(y, z)$-plane. Here, a good fit can only be obtained with same signs of $g_y$ and $g_z$ (dotted lines). The best value for $\xi$ is 0.22, suggesting that DNP is more efficient for $\mathbf{B}$ along $y$ than along $x$. The absolute signs of the g-tensor elements are obtained using the same arguments given for the 135 K data. We find that $g_z$ and $g_y$ are both negative, whereas $g_x$ must be positive. The data is summarized in Tab. I.

With this model, we can predict how the nuclear polarization would behave with an isotropic g-factor. The dash-dotted lines in Fig. 4 shows data obtained assuming $g_{x,y} = g_z$. For easier comparision, the nuclear contribution to $\Omega_L$ was added onto an angle-dependent Zeeman background $\mathbf{\Omega}$ corresponding to the respective anisotropic g-tensor. The data shows that for $\mathbf{B}$ in the $(x, z)$-plane, the anisotropy enhances nuclear polarization for $-10° < \alpha < 7°$, whereas for $\mathbf{B}$ in the $(y, z)$-plane, the enhancement extends from 10° to all negative angles. The increased nuclear polarization is due to the finite $\mathbf{S}_\parallel$, and is advantageous for Larmor magnetometry in the quasi-transverse geometry.

In the following, we describe how the nuclear field, $B_n$, can be obtained from the measurement. Using Eq. (4), we rewrite the definition of the nuclear field,[14] $\mathbf{B_n} = A\langle \mathbf{I} \rangle / g\mu_B$, as $\mathbf{B_n} = \hbar(\mathbf{\Omega}_{tot}-\mathbf{\Omega})/g\mu_B$. From the experiment, we obtain $\Omega_L = |\mathbf{\Omega}_{tot}|$ (symbols in Fig. 4) and $\Omega = |\mathbf{\Omega}|$ (dashed line in Fig. 4). Because of the anisotropic g-factor, $\mathbf{\Omega}_{tot}$ and $\mathbf{\Omega}$ are not collinear, and $B_n$ is generally not proportional to the difference of $\Omega_{tot}$ and $\Omega$. However, for small $\alpha$, and if the nuclear field enhances $\Omega_L$, i.e. when $\mathbf{\Omega}_{tot}$ and $\mathbf{\Omega}$ point approximately into the same direction, $\Omega_{tot} \approx \Omega + \Omega_n$ and $B_n \approx B(\Omega_L - \Omega)/\Omega$ is a good approximation. For GaAs, it was predicted[13] that the maximum nuclear field is 5.3 T for 100% nuclear polarization. This value scales with $g_{GaAs}/g$ for g-factors different from the bulk GaAs value of $g_{GaAs} = -0.44$. This leads to comparably large nuclear fields for low electron g-factors. As an example, we measure $\Omega_L = 50$ GHz for $\alpha = -20°$, as shown in Fig. 4(b). The Zeeman contribution is $\Omega \approx 23$ GHz, and $B_n \approx B(\Omega_L - \Omega)/\Omega \approx 4.7$ T. If one considers that $\mathbf{\Omega}$ and $\mathbf{\Omega_n}$ are at an angle of $\gamma - \alpha \approx 35°$, one gets $B_n \approx 5.1$ T. Since at this angle, the absolute value of the g-factor is 0.067, the maximum nuclear field amounts to 35 T, and the measured $B_n$ corresponds to a nuclear polarization of 15%.

## IV. ALL-OPTICAL NMR

In this section, we study the resonant depolarization of nuclear spin induced by the periodicity of laser pulse arrivals. The nuclear gyromagnetic ratios $\gamma$ for the three different isotopes in GaAs are $\gamma(^{71}Ga) = 8.158 \cdot 10^7$ rad/Ts, $\gamma(^{69}Ga) = 6.421 \cdot 10^7$ rad/Ts, and $\gamma(^{75}As) = 4.578 \cdot 10^7$ rad/Ts. The pulse repetition rate is 76 MHz. The light pulses interact indirectly with the nuclei via the electron system. In Refs. 19,9, it was proposed that the contact hyperfine interaction of electron spins with nuclear spins might lead to tipping of nuclear spins. As was shown in Ref. 10, pulse repetition rates at twice the nuclear Larmor frequency $\omega = \gamma B$ can induce resonant transitions of $\Delta m = 2$ within the nuclear spin levels with angular quantum number $m$. Such transitions at $2\omega$ are not allowed by selection rules for magnetic dipole coupling, i.e. they are inconsistent with contact hyperfine coupling. We rule out occurrence of multi-spin transitions as reported in Ref. 32, because we do not observe any resonances at sum frequencies of different isotopes or at 3ω. Furthermore, the proposed effect was observed at small fields of a few Gauss and decreases with $1/B^2$, ruling out its occurrence at higher fields. However, the interaction of the nuclear electric quadrupole moment with modulated electric field gradients can induce transitions at 2ω excitation, as was experimentally observed in bulk GaAs by modulating an electric field across the semiconductor.[33] In the case of all-optical NMR, electric-field gradients might be modulated by periodically injected charges,[10,34] in contrast to the hyperfine scenario. We are able to experimentally separate the two mechanisms and find that the charge of the electrons, not the spin, is responsible for nuclear resonances at 2ω, and that both mechanisms play a role at ω excitations.

Figure 5(a) shows the nuclear fields obtained by time-resolved FR scans continuously taken while $B$ sweeps slowly (1 mT/min) across the resonances of the $^{69}$Ga isotope. The laser repetition rate, i.e. the excitation frequency, is fixed at 76.000 MHz. The average laser pump power is 1.2 mW, and the sample is tilted by $\alpha = 5°$. The helicity of the pump pulses is chosen such that the nuclear field increases the electron precession frequency. The ω and the 2ω resonances occur at $B_0 = 7.44$ T and 3.72 T, respectively. The resonance is plotted as a function of $\Delta B = B - B_0$. We find a background nuclear polarization of $B_n^0 = 8.0$ T around the ω excitation, and $B_n^0 = 8.8$ T around the 2ω excitation [dashed lines in Fig. 5(a)]. On resonance, $B_n$ decreases due to nuclear depolarization. Clearly resolved are two dips in $B_n$ at 3.72 T and five dips at 7.44 T. The number of dips and their relative field positions agrees with the picture of $\Delta m = 1$ and $\Delta m = 2$ transitions within the spin-3/2 nuclear levels, induced by the 76 MHz pulse train and its second-harmonic 152 MHz component,[10] as shown in Fig. 5(b). The maximum relative depolarization for the $\omega$ excitation is 14%, whereas the $2\omega$ excitation leads to a depolarization by 12%.

In the following, we make use of sidebands that occur in the excitation spectrum when the pump pulse train is modulated. Information about the resonant tipping mechanism is obtained by comparing the resonance curves under modulation of amplitude or helicity of the pump pulses. This varies either the charge or the spin of the periodically injected electrons, affecting the resonance only if the respective quantity induces nuclear tipping. If the pump beam is mechanically chopped at frequency $\nu_{ch}$, the excitation frequency spectrum contains sidebands at integer multiples of $\nu_{ch}$ [inset of Fig. 6(a)]. We observe such sidebands in the resonance curve only for $\nu_{ch} > 3$ kHz. For smaller $\nu_{ch}$, the spectral distance of the sidebands is smaller than the resonance line width, which therefore is on the order of 3 kHz, in agreement with values found earlier.[10] Figure 6(a) shows the resonance at 7.44 T with $\nu_{ch} = 4.1$ kHz and 6.1 kHz. The average pump-beam intensity is 1.4 mW and $\alpha = 10°$. Instead of the more time-consuming method of extracting the nuclear field across the resonance as shown in Fig. 5, here we simply measure FR at fixed time-delay $\Delta t$. A change in $B_n$ leads to a change of the spin phase at the given time delay, which is reflected in the FR signal. We choose $\Delta t$ close to zero-crossings of the FR oscillations, for maximum sensitivity to changes in the spin phase. An increase of the FR signal shown in Fig. 6(a) corresponds to a decrease of $B_n$. The quadrupolar features are not visible, and periodic oscillations appear instead. A Fourier transform of FR vs. $B$ reveals that the field periods $\Delta B$ correspond to frequencies $\gamma\Delta B/2\pi$ matching $\nu_{ch}$ and $2\nu_{ch}$ [Fig. 6(b)], indicating that the even and odd harmonics in the sideband spectrum are weighted differently, as depicted in the insert of Fig. 6(a). This is expected for a symmetric modulation that contains mainly odd higher harmonics.

In order to better resolve individual sideband peaks, we modulate the amplitude at higher frequencies by using a photoelastic modulator instead of a mechanical chopper. A sequence of a photoelastic modulator (frequency 50.1 kHz, $\lambda/2$ retardance), a linear polarizer, and a $\lambda/4$ retarder is used to generate circularly polarized pump pulses amplitude-modulated at 100.2 kHz. Curve *A* in Fig. 7(a) shows the resonance spectrum obtained for the $^{69}$Ga isotope excited at $\omega$. The $\Delta m = 1$ triplet is indicated with a threefold bracket above the curve, and the $\Delta m = 2$ doublet with twofold brackets below the curve. The triplet peaks are repeated at field intervals corresponding to 100.2 kHz. Because the $\Delta m = 2$ transitions at 7.44 T are excited with the second harmonic component of the laser pulse train, the sideband separation is 50.1 kHz instead of 100.2 kHz. Small peaks occur also at field positions that would be consistent with 50.1 kHz modulation of $\Delta m = 1$ transitions. They are probably due to a slightly mistuned $\lambda/2$ retarder. An accidental coincidence of the quadrupole splitting with one third of the modulation frequency leads to an overlap of displaced doublet peaks with triplet peaks. This makes a quantitative analysis of the peaks difficult.

Alternatively, we can modulate the helicity of the circularly polarized pump pulses at 50.1 kHz using the same photoelastic modulator set at $\lambda/4$ retardance without the linear polarizer. This leads to periodically injected electrons with modulated spin orientation. Although much smaller in amplitude, DNP still takes place with this set-up, contrary to the general belief of Ref. 14. We enhance the nuclear polarization further by slightly detuning the balance between left- and right-circular polarization. Curve *B* in Fig. 7(a) shows the resonance spectrum. If the electron spin is not connected to the nuclear depolarization mechanism, then no sidebands should be observed in the resonance spectrum. However, we clearly resolve sidebands, suggesting that spin is important for the resonant depolarization of nuclear spin at $\omega$ excitation. For comparison, curve *C* shows the resonance where the pump is chopped at 2 kHz, showing no sidebands. However, the linewidths of the quadrupolar-split resonances are significantly larger here, indicating a broadening from the low-frequency sidebands in the excitation spectrum.

Figure 7(b) shows similar measurements as Fig. 7(a), but for the $2\omega$ excitation of $^{75}$As. Here, sideband peaks occur at 100.2 kHz separation for amplitude modulation (curve *A*), but only very weak sideband peaks are visible for spin modulation (curve *B*). Unintentional amplitude modulation of the laser pulse train right before the cryostat window is measured to be smaller than $10^{-3}$. However, we can not exclude small charge modulations inside the sample. As a reference, Curve *C* shows the broadened resonance peak for mechanically chopped pump pulses.

A more direct way to study the tipping mechanism is to compare the resonance induced by a circularly polarized pump beam with that from a linearly polarized one. In the first case, the polarized electron spin allows for periodic contact hyperfine interaction with the nuclear spin, whereas in the second case, up and down spins are populated equally, eliminating efficient hyperfine coupling. Because Larmor magnetometry relies on circularly polarized pump pulses injecting spin-polarized carriers, we cannot tune the pump to linear polarization. Therefore, we focus a second pump beam onto the sample, whose pulse repetition rate, energy and polarization can be tuned independently.

In Fig. 8(a), data is shown for the $^{69}$Ga resonance excited at $\omega$. The first and second pump pulse trains have repetition rates of 76.000 MHz and 75.620 MHz, respectively. Both beams are tuned to 1.572 eV. The first pump beam is always circularly polarized and is chopped at 2.0 kHz, the second at 1.8 kHz. The measured electron-spin precession frequencies $\Omega_L$ are transformed to nuclear fields $B_n$ and plotted as relative nuclear polarization $B_n / B_n^0$, where $B_n^0$ is the nuclear field slightly off resonance. The first beam induces a resonance at $\Delta B = 0$ T. The second-pump resonance occurs at lower fields, and clearly depends on the beam's polarization. Circularly polarized pulses (circles) depolarize the nuclei by 10%, whereas the depolarization for linearly polarized pulses (diamonds) is only 3%. Note that although the overall degree of DNP is smaller with the second pump linearly polarized, the relative depth of the first-pump resonance does not depend on the second pump's polarization. The quadrupolar features in the second-pump resonance are barely visible for circularly polarized pulses. If the $\Delta m = 2$ doublet is insensitive to the electron spin polarization, as suggested by the sideband experiments described above, then the decrease in resonance depth is entirely due to a decrease of the $\Delta m = 1$ triplet, and the $\Delta m = 2$ doublet should gain in relative strength. However, the data is not conclusive enough to support this idea.

Figure 8(b) shows data of the $^{75}$As resonance at $2\omega$ excitation. Here, the repetition rate of the second pump beam is detuned by 290 kHz. The data shows two doublet structures. The one around $\Delta B = 0$ T originates from the two $\Delta m = 2$ resonances induced by the first pulse train, whereas the second pulse train induces a resonance doublet about 20 mT lower in field, corresponding to the detuning in repetition rate of 290 kHz. Circularly and linearly polarized pump beams induce resonances of the same depth (about 17%), indicating that the spin degree of freedom of the injected carriers is not relevant for the $\Delta m = 2$ tipping process. Also shown is data for the second pump beam detuned in energy to 1.540 eV, which is below the absorption edge of the QW. The second resonance doublet completely disappears. The tipping process is therefore not directly related to the electromagnetic field of the laser pulses. This indicates that modulating the carrier properties is essential for mediating the interaction of the laser pulse train with the nuclear spins.

## V. CONCLUSIONS

The technique of Larmor magnetometry relies on injection of electron spin transverse to an applied magnetic field and subsequent measurement of its precession. On the other hand, the electron spin should be oriented along the field direction in order to achieve strong DNP. As demonstrated in this paper, efficient nuclear polarization is possible in a quasi-transverse geometry by using a QW sample with a strongly anisotropic g factor. This tilts the precession axis away from **B**, leading to a nonprecessing electron-spin component along **B**. While convenient, we note that anisotropic g factors are not necessary for this technique, as nuclear polarization – though weaker – should also occur for an isotropic g factor. DNP can in principle be enhanced further with an additional pump beam oriented along the field direction. We determined the three components of the strongly anisotropic electron g-tensor, including its signs at 5 K, and studied the direction of the precession axis and of the injected electron spin. At liquid He temperatures, the measured precession frequency strongly depends on the nuclear spin polarization, allowing for sensitive nuclear magnetometry. We investigated the mechanisms of all-optical NMR, where the nuclear spins are resonantly manipulated by a laser pulse train. Resonances excited at $2\omega$ are solely attributed to interaction of the electron charge with the nuclear quadrupole moments, whereas resonances at $\omega$ are due to both hyperfine coupling of the periodically injected electron spin to the nuclear spin, and electric quadrupolar coupling. If the tipping fields prove to be large and homogenous enough, pulsed techniques within the all-optical NMR scheme might enable coherent manipulation of local nuclear spins driving simultaneously $\Delta m = 1$ and $\Delta m = 2$ transitions.

## ACKNOWLEDGEMENTS

We thank M. E. Flatté and J. M. Kikkawa for helpful discussions, and J.A. Gupta and R. J. Epstein for their careful reading of the manuscript. Support from the ARO DAAG55-98-1-0366, DARPA/ONR N00014-99-1096, and NSF DMR-0071888 is acknowledged. The work at Tohoku University was supported by the Ministry of Education, Japan (#12305001) and by the Japan Society for the Promotion of Science (JSPS-RFTF97P00202).

Table I. Measured components of the electron g-tensor in a 7.5 nm wide (110) GaAs QW at 135 K and 5 K. The negative sign of $g_z$ relies on the assumption $g_z < g_x, g_y$.

|       | $g_x$ <br> $x \parallel [\bar{1}10]$ | $g_y$ <br> $y \parallel [001]$ | $g_z$ <br> $z \parallel [110]$ |
|-------|---------------------------|-------------------------|-------------------------|
| 135 K | (±) 0.0376 ± 0.0004       | (±) 0.0184 ± 0.0009     | -0.142 ± 0.001          |
| 5 K   | 0.031 ± 0.005             | -0.041 ± 0.005          | -0.16 ± 0.02            |

**FIGURE CAPTIONS**

Fig. 1. (a) The sample is mounted so that either $x \parallel [\bar{1}10]$ or $y \parallel [001]$ can be tilted by an angle $\alpha$ with respect to the applied field, **B**. Pump and probe beams are perpendicular to **B**. (b) The injected electron spin **S** precesses about an axis **Ω** whose direction depends on $\alpha$. Time-resolved electron spin precession at $T = 135$ K for two angles $\alpha = 0°$ and $25°$, and a field of 4 T applied in the $(x, z)$-plane is shown in (c). Circles are measured FR data. The angle-dependent oscillation frequency arises from an anisotropic electron g-factor. The data is fit (solid lines) by an exponentially decaying harmonic oscillation added to a second, non-precessing exponential decay.

Fig. 2. (a) The electron g factor vs. $\alpha$ is obtained from the oscillation frequency of fits to FR data as shown in Fig. 1(c). Filled circles and squares are for the two orientations of the in-plane sample directions. The solid line fits the angle-dependence of $g$, giving the two components $g_x$ and $g_z$ (circles) or $g_y$ and $g_z$ (squares). (b) The amplitudes $a_1$ of the non-precessing FR component are compared to a geometric model assuming spin-injection along the refracted pump beam direction, where same (solid line) and opposite (dotted line) signs for $g_{x,y}$ and $g_z$ are chosen. The dashed lines represent data assuming spin orientation strictly along the QW confinement axis.

Fig. 3. Measurement of the apparent electron g factor at 5 K, as obtained from fits to time-resolved FR at $B = 4$ T for B applied in the $(x, z)$-plane (a) and $(y, z)$-plane (b). Open diamonds are for right-circularly polarized pump pulses, filled symbols for left-circular excitation at two different laser intensities. The electron precession frequency is largely affected by hyperfine coupling to dynamically polarized nuclear spins. Lowering the pump intensity (solid squares) decreases nuclear polarization and reveals the bare electron g factor, which is obtained by fitting the lowest-intensity data to the angle-dependence of an anisotropic g-factor, thus obtaining the three components $|g_x|$, $|g_y|$ and $|g_z|$ at 5 K.

Fig. 4. Calculated angle-dependence of the electron-spin precession frequency $\Omega_L$ including the nuclear field. The solid lines are for different signs of the in- and out-of-plane components of $\hat{g}$ with absolute values determined by the fits taken from Fig. 3. The dotted line assumes same signs of the two components. The dashed line shows the Zeeman frequency alone. The dash-dotted line represents data for nuclear polarization calculated assuming an isotropic $\hat{g}$ and added onto the anisotropic Zeeman frequency. The external field **B** was applied in the $(x, z)$-plane (a) or in the $(y, z)$-plane (b). The open diamonds represent the measured data already shown in Fig. 3. The fits indicate that $g_x$ and $g_z$ have different signs, whereas $g_y$ and $g_z$ have the same sign (which is negative). The fit parameter $\xi$ represents the maximum relative nuclear polarization and is weaker in (a).

Fig. 5. Spectra of all-optical nuclear resonance induced by the pump laser pulse-train which has a repetition rate of 76 MHz (a). The nuclear field $B_n$ acting on electron spins is measured at the $^{69}$Ga resonances at $\omega$ and $2\omega$ excitation, occurring at $B_0 = 7.44$ T and 3.72 T, respectively. The resonances are plotted vs. $\Delta B = B - B_0$, and show signatures of quadrupolar splittings. Arrows indicate triplet and doublet structures, corresponding to transitions with changes in angular quantum number $\Delta m = 1$ and $\Delta m = 2$. Comparing the depth of resonance with the background polarization (dashed lines), gives nuclear depolarizations of 12% for $2\omega$, and 14% for $\omega$ excitation. In (b), the transitions within the nuclear spin levels at $\omega$ and $2\omega$ excitation are shown. Solid arrows indicate the $\Delta m = 1$ and $\Delta m = 2$ transitions induced by the excitation at 76 MHz, whereas the dashed arrows show $\Delta m = 2$ transitions induced by the second-harmonic component at 152 MHz.

Fig. 6. (a) Spectra of the $^{69}$Ga resonances at 7.44 T, obtained by measuring the FR signal at fixed time delay $\Delta t = 320$ ps and $\alpha = 10°$. The pulse train is modulated at 6.1 kHz and 4.1 kHz, respectively, leading to sidebands in the excitation spectra, as depicted in the inset. The resonance is broadened, quadrupolar features are not resolved and oscillations periodic in $B$ appear. (b) A Fourier transformation of the resonance reveals that the periods correspond to spacings in the sideband spectra of one and two times the modulation frequeny $\nu_{ch}$.

Fig. 7. (a) Resonance of $^{69}$Ga at 7.44 T. The excitation is amplitude-modulated with 100.2 kHz in (A). The fundamental resonance consists of five peaks, corresponding to a $\Delta m = 1$ triplet (threefold brackets) and a $\Delta m = 2$ doublet (twofold brackets). The triplet recurs periodically displaced by fields corresponding to 100.2 kHz, whereas the doublets are periodically displaced by 50.1 kHz. In (B), the helicity of circular polarization is modulated at 50.1 kHz. The data can be explained by considering periodically displaced triplets and only the central doublet. As a reference, (C) shows the signal where the pump-pulse amplitude is modulated at a small frequency of 2 kHz. The appearance of sidebands in (B) demonstrates that the resonance mechanism is sensitive to the spin of the injected electrons. (b) Similar data as in (a) for the $^{75}$As resonance at $2\omega$ excitation, occurring at 5.21 T. The absence of pronounced sidebands in the spectrum (B) suggests that no spin mechanism accounts for the resonant tipping at $2\omega$ excitation.

Fig. 8. (a) Resonance of $^{69}$Ga at $\omega$ excitation for two circularly polarized laser pump trains with repetition rates of 76.000 MHz and 75.620 MHz, respectively, simultaneously exciting electrons in the quantum well. The first pulse train is circularly polarized and is used for Larmor magnetometry of the nuclear fields. It induces a resonance at $\Delta B = 0$ T. The second pulse train leads to an additional resonance at lower fields. When its polarization is tuned from circular (diamonds) to linear (open circles), the depth of this second resonance significantly decreases. This indicates that the depolarization mechanism is sensitive on the spin-polarization of the periodically injected electrons. (b) The form and depth of the $^{75}$As $2\omega$ resonance shows no dependence on polarization. Here, the second pump is detuned in repetition rate by 290 kHz, leading to a Δ*m* = 2 doublet displaced by about 20 mT (arrow). If the energy is tuned below the absorption edge of the QW to 1.540 eV (crosses), this resonance disappears.

[28] No in-plane anisotropy is expected for a symmetric (001) QW because its symmetry ($D_{2d}$) is higher than in a (110) QW ($C_{2v}$). However, an anisotropic electron g-factor was predicted in a biased (001) QW by V. K. Kalevich and V. L. Korenev [JETP Lett. **57**,

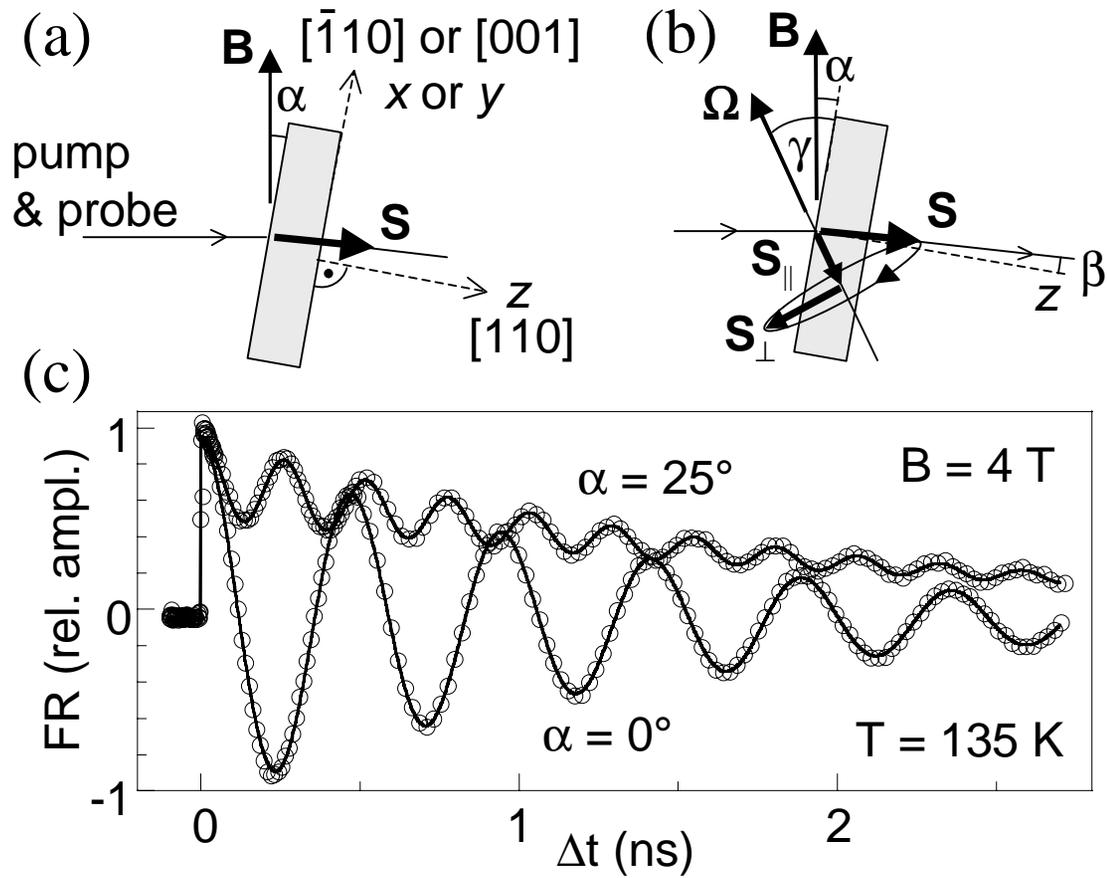

Salis, et al, Fig. 1

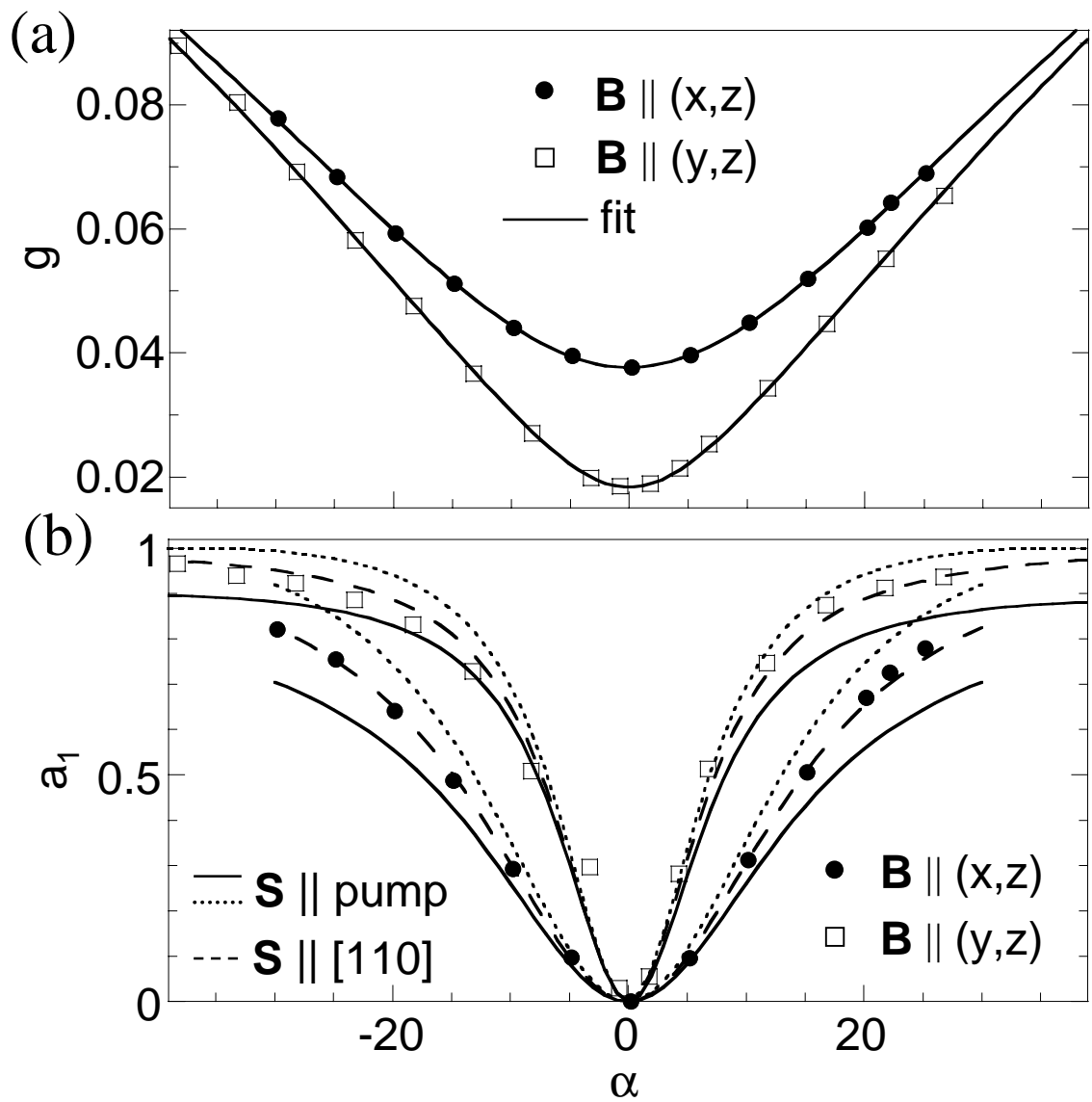

Salis, et al, Fig. 2

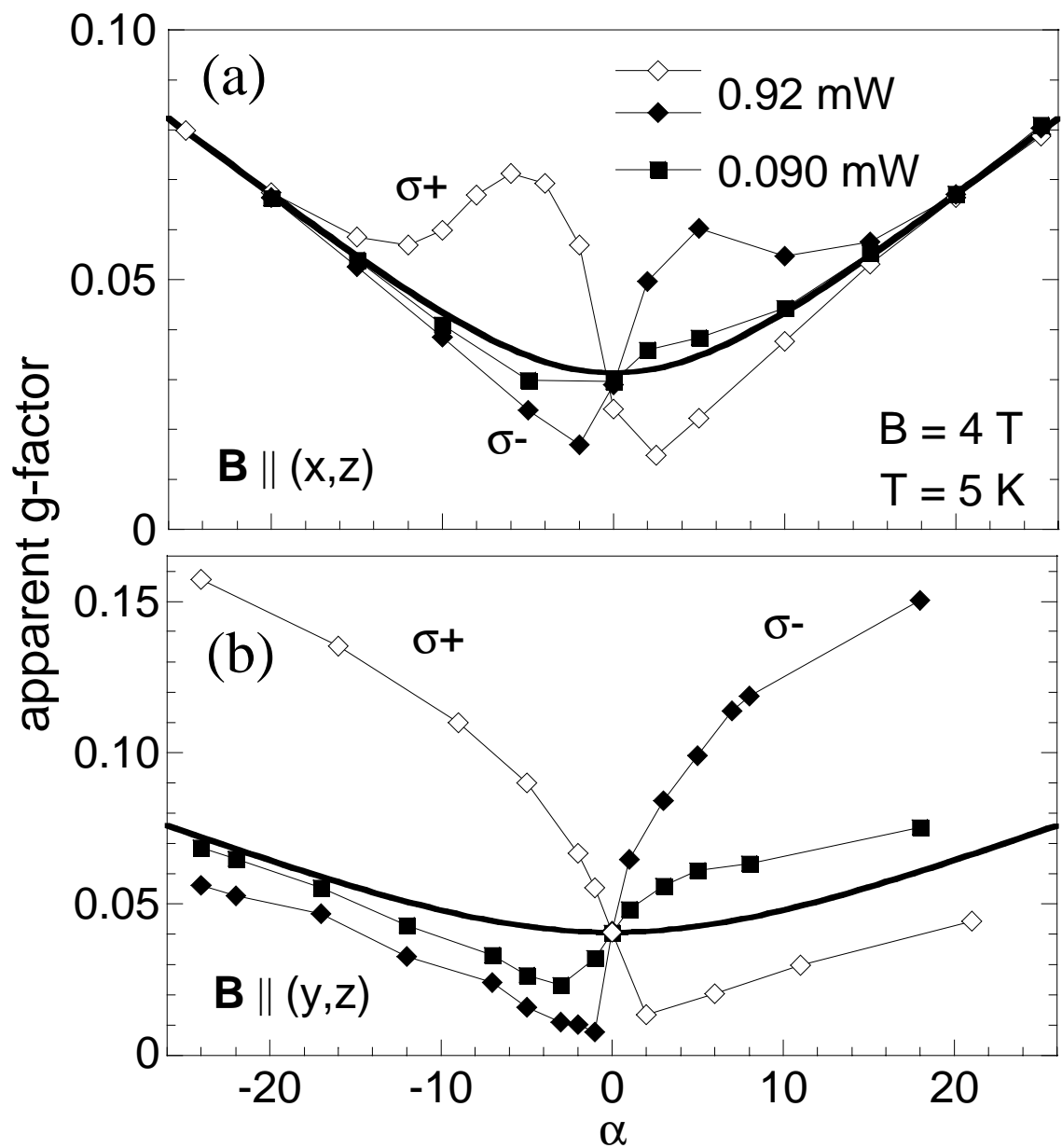

Salis, et al, Fig. 3

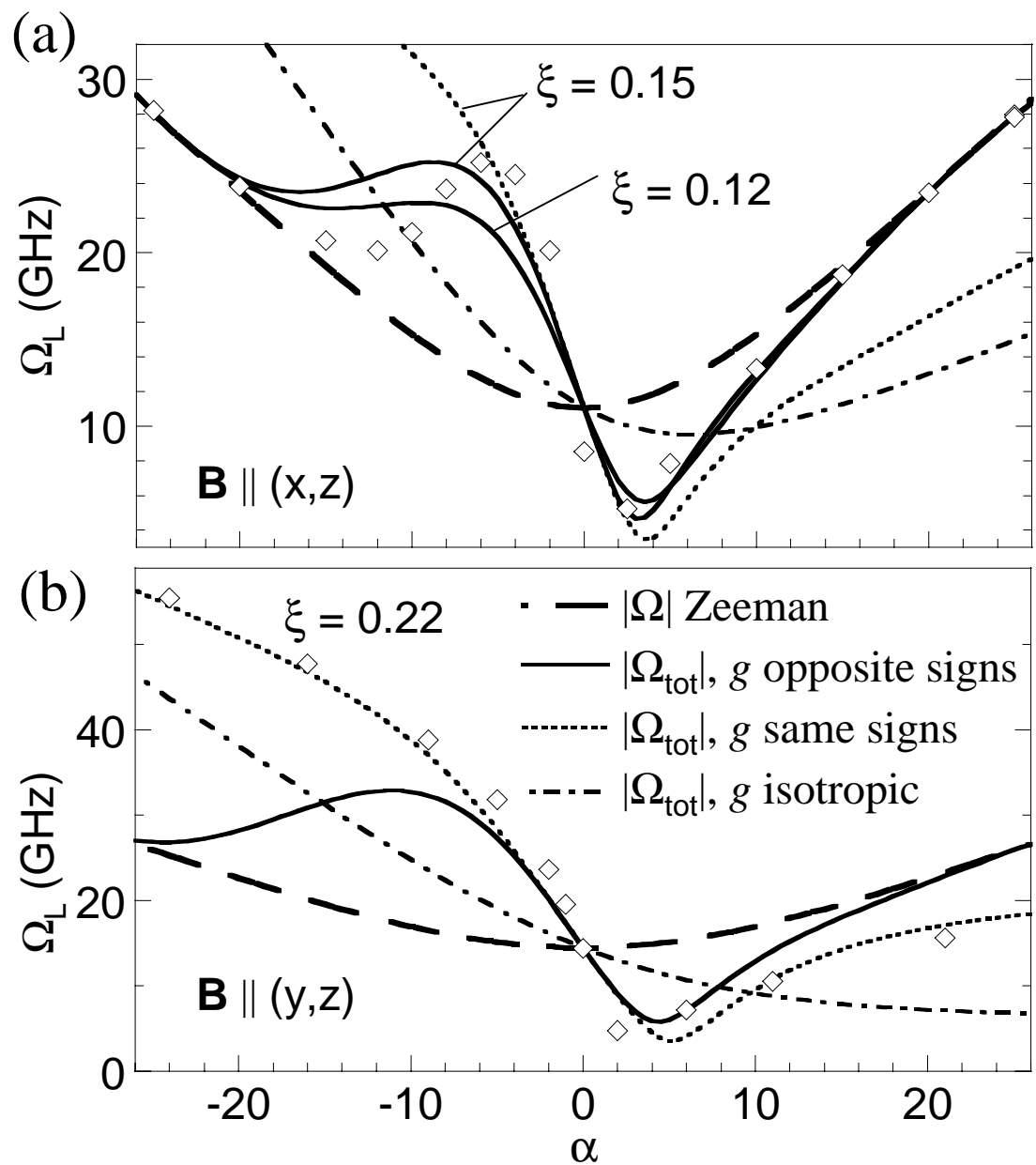

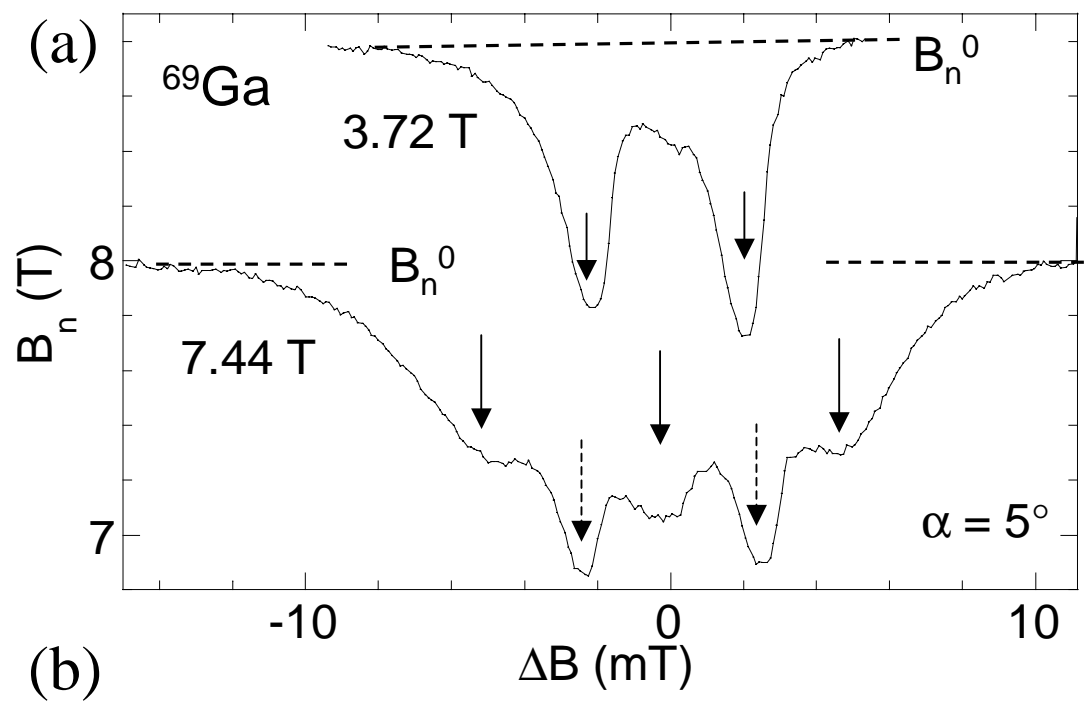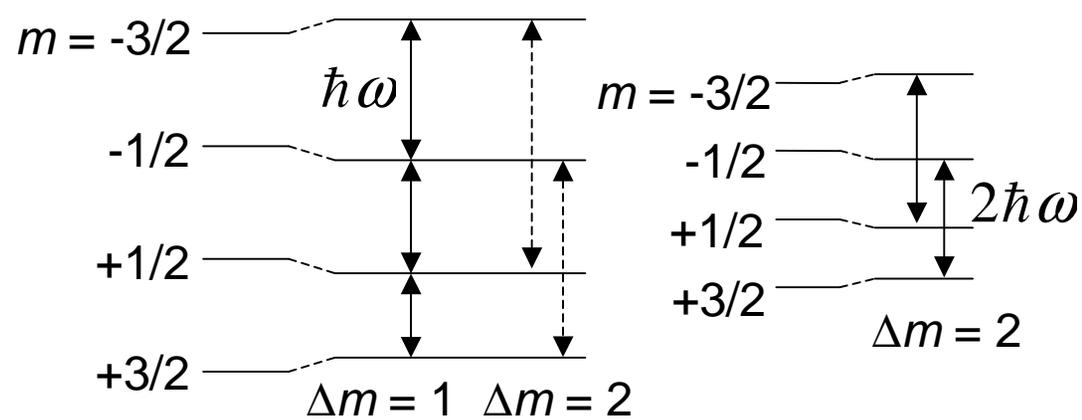

Salis, et al, Fig. 5

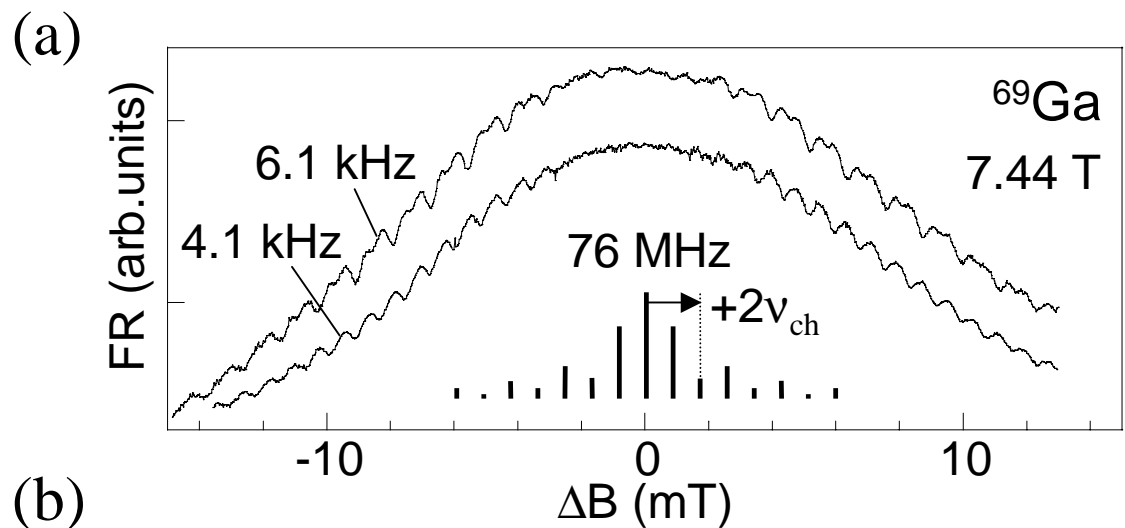
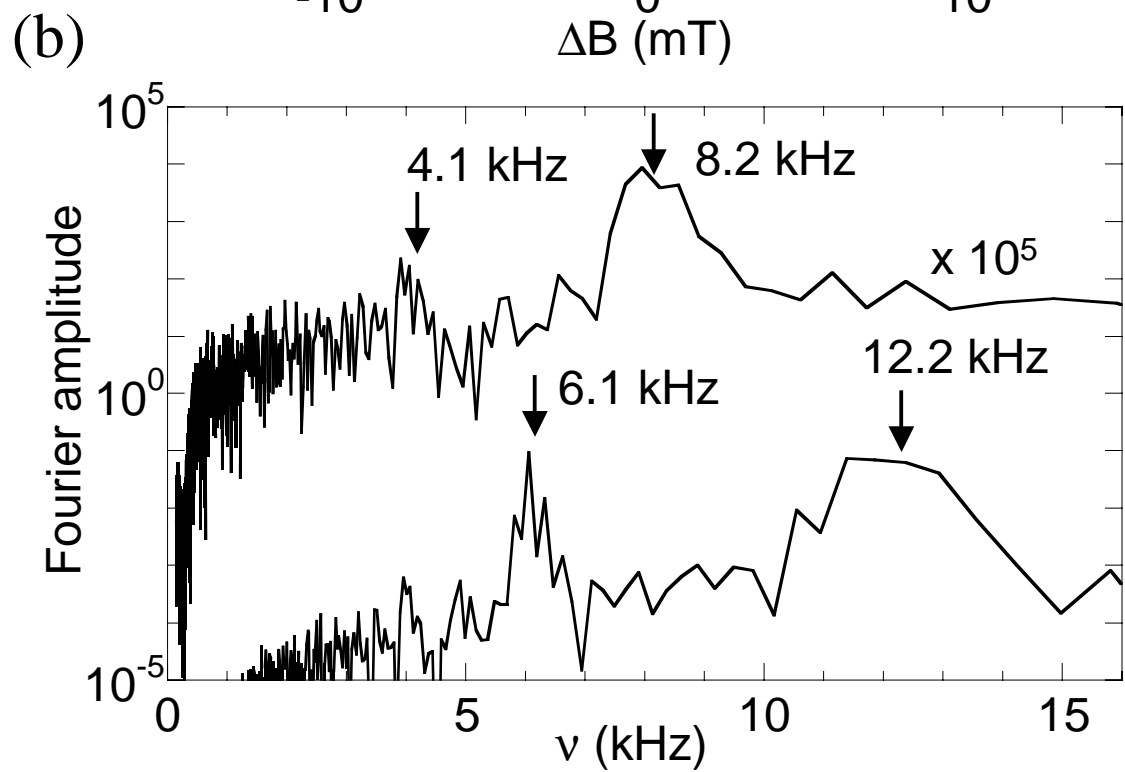

Salis, et al, Fig. 6

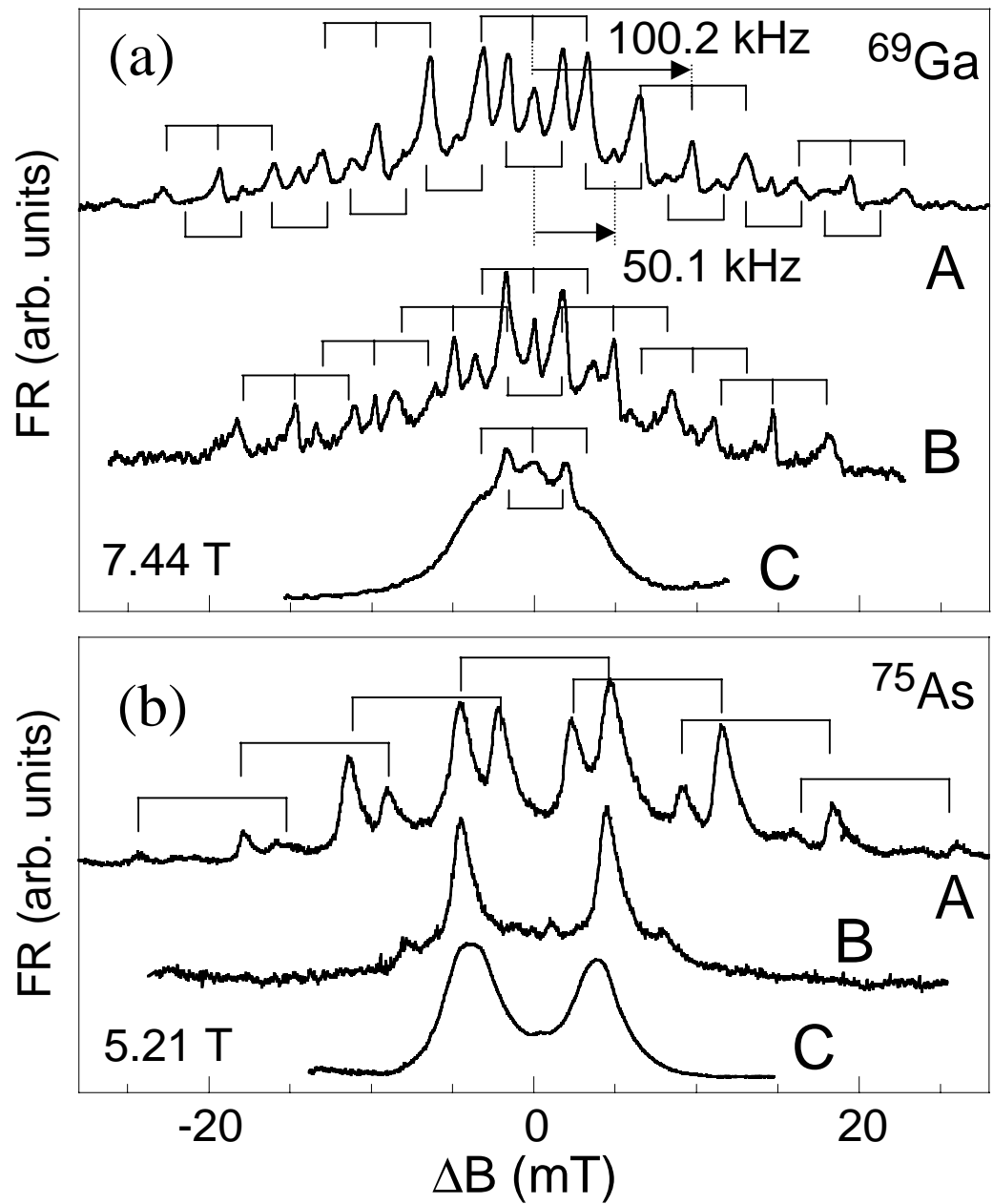

Salis, et al, Fig. 7

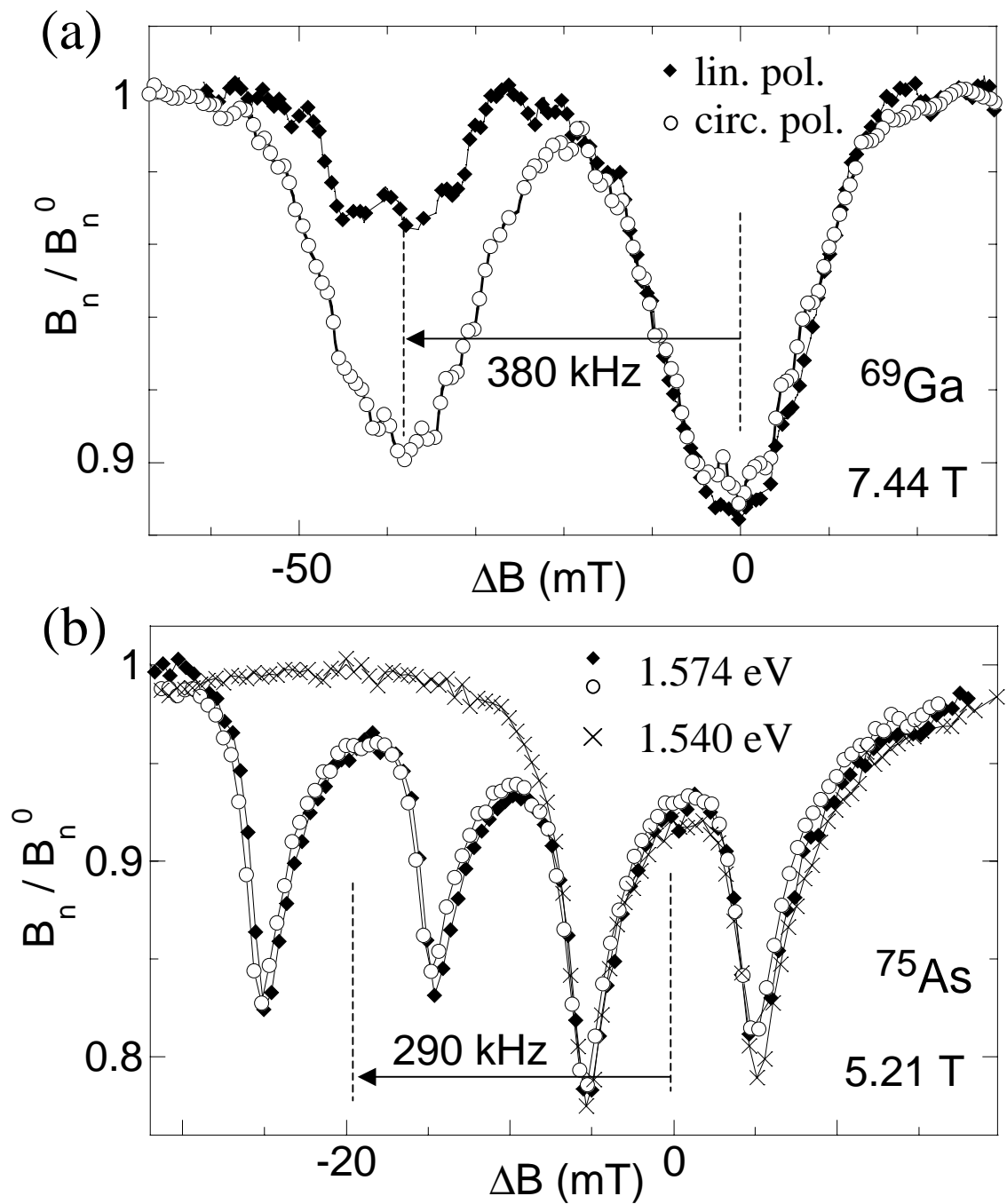

Salis, et al, Fig. 8